\documentclass[12pt]{iopart}

\usepackage{graphicx}

\begin{document}
\title[Remarks on Newton's Second Law]
{Remarks on Newton's Second Law 
for Variable Mass Systems}

\author{K. Nakayama}

\address{Department of Physics and Astronomy, University of Georgia,
Athens, GA 30602, USA}
\ead{nakayama@uga.edu}

\begin{abstract}
Misinterpretations of Newton's second law for variable mass systems found in the literature are addressed. In particular, 
it is shown that Newton's second law in the form $\vec{F} = \dot{\vec{p}}$ is valid for variable mass systems in general, contrary to the claims by some authors that it is not applicable to such systems in general. In addition, Newton's second law in the form $\vec{F} = m\, \dot{\vec{v}}$ -- commonly regarded as valid only for constant mass systems -- is shown to be valid also for variable mass systems. Furthermore, it is argued that $\vec{F} = m\, \dot{\vec{v}}$ may be considered as the fundamental law, alternatively to $\vec{F} = \dot{\vec{p}}$.
 The present work should be of direct relevance to both instructors and students who are engaged in teaching and/or learning classical mechanics in general and variable mass systems in particular at a middle- to upper-level university physics. 
  \end{abstract}


\submitto{\EJP}
\maketitle

\normalsize

\section{Introduction}

In most of the standard textbooks \cite{Goldstein:2000,Thornton:2003} the Newton's second law in the form
\begin{equation}
\vec{F} = \frac{d \vec{p}}{dt} \equiv \dot{\vec{p}} 
\label{eq:1a}
\end{equation}
is considered to be valid for all systems and when the system's mass is constant, it reduces to
\begin{equation}
\vec{F} =  m \frac{d\vec{v}}{dt}  \equiv m \dot{\vec{v}} \ .
\label{eq:1b}
\end{equation}
In the above equations, $\vec{p}$ stands for the (linear) momentum defined as $\vec{p} \equiv m \vec{v}$, with $m$ denoting the mass of the system and, $\vec{v}$, its velocity.

Application of Newton's second law to variable mass systems has been a source of some confusions and disagreements in the past and some authors \cite{Thorpe:1962,Arons:1964,Tiersten:1969,Siegel:1972,Kleppner:1973,Plastino:1992,deSousa:2004} have addressed this issue specifically. 
They all conclude that Newton's second law in the form given by Eq.~(\ref{eq:1a}) does not hold for variable mass systems in general, and that, it is no more general than Eq.~(\ref{eq:1b}). More specifically, the authors of Refs.~\cite{Thorpe:1962,Arons:1964,Tiersten:1969,Siegel:1972,deSousa:2004} state that  Eq.~(\ref{eq:1a}) applies only to two special situations, while the authors of Refs.~\cite{Kleppner:1973,Plastino:1992} simply state that Eq.~(\ref{eq:1a}) is not valid for variable mass systems. 
The major reasons for the authors 
arriving at this conclusion are twofold. One is that, for variable mass systems, it is well known that  the correct equation-of-motion describing such systems is \cite{Sommerfeld:1952}
\begin{equation}
\vec{F} = \frac{d \vec{p}}{dt} - \frac{dm}{dt} \vec{u} \equiv \dot{\vec{p}} - \dot{m} \vec{u} \ ,
\label{eq:2}
\end{equation}
as derived from momentum conservation theorem; it can be derived from other approaches as well \cite{Thorpe:1962,Tiersten:1969,deSousa:2004}.
In the above equation, $\vec{u}$ denotes the velocity of the amount of mass of the system being changed at the rate $\dot{m}$. 
The other reason, as pointed out in Ref.~\cite{Plastino:1992}, is that Eq.~(\ref{eq:1a}) apparently leads to the violation of the Galilean principle of relativity for, in a primed inertial frame moving with a constant velocity $\vec{v}_0$ with respect to the unprimed inertial frame,  $\vec{v}\,' = \vec{v} + \vec{v}_0$, Eq.~(\ref{eq:1a}) leads to
\begin{equation}
\vec{F}\,' = \dot{\vec{p}}\,' = \frac{d}{dt}(m\vec{v}\,') = m\, \dot{\vec{v}} + \dot{m}\, \vec{v} + \dot{m}\, \vec{v}_0 \ .
\label{eq:3}
\end{equation}

\vskip 0.5cm
In the following -- contrary to the claims found in the literature \cite{Thorpe:1962,Arons:1964,Tiersten:1969,Siegel:1972,Kleppner:1973,Plastino:1992,deSousa:2004} -- we show that Eq.~(\ref{eq:1a}) is valid for variable mass systems in general, and that, no violation of the Galilei's principle of relativity exists. 
 In addition, Newton's second law in form (\ref{eq:1b})  -- commonly regarded as valid only for constant mass systems -- is shown to be valid for variable mass systems as well. Also, it is argued that Eq.~(\ref{eq:1b}) may be considered as the fundamental law, alternatively to Eq.~(\ref{eq:1a}) or to Eq.~(\ref{eq:2}) for that matter.  
 It should be made clear that our criticism to Refs.~\cite{Thorpe:1962,Arons:1964,Tiersten:1969,Siegel:1972,Kleppner:1973,Plastino:1992,deSousa:2004} is not about the derivations/calculations presented there
 but to the statement that Eq.~(\ref{eq:1a}) is not valid in general for variable mass systems.   
Even though some of the aspects in the following have been discussed in the past, we include them here as an integral part in addressing the issue at hand.

The present work should be of direct relevance to both instructors and students alike who are involved in teaching and/or learning classical mechanics in general and variable mass systems in particular at a middle- to upper-level university physics. By clarifying the existing issues on variable mass systems in the literature, this article should help avoid potential confusions and misinterpretations by those engaged in teaching and/or learning this topic. Furthermore, it offers a choice to treat the 
variable mass systems on same footing with the constant mass systems, a feature that greatly facilitates solving problems 
involving variable mass systems in practice.

\section{Analysis}\label{sec:analysis}

\vskip 0.2cm
We begin by focusing our attention on the term $\dot{m}\, \vec{u}$ appearing in Eq.~(\ref{eq:2}).
The key issue here is to realize that the system in which the mass is changing cannot be a \textit{closed (or isolated) system} in general
\footnote{Closed system is a system where the total external force is zero.}
and that $\dot{m}\, \vec{u}$ is a part of the external force. It accounts for the 
rate at which the momentum is being transferred to or removed from the system.
It is this transferred momentum that is responsible, e.g., for the recoil effect. In fact, when a gun is fired, the expelled bullet 
exerts a force on the gun which causes it to recoil or when the rocket expels the fuel in the form of a exhaust,  the latter exerts a force on the rocket providing the thrust to the rocket. The fact that $\dot{m}\, \vec{u}$ belongs to the external force 
is further discussed in \ref{app:1}.  
Accordingly, we may incorporate $\dot{m}\,\vec{u}$ into $\vec{F}$, so that Eq.~(\ref{eq:2}) 
becomes of the form (\ref{eq:1a})  
with $\vec{F}$ there being the total external force including
\begin{equation}
\vec{F}^{\rm mass}_u \equiv \dot{m}\,\vec{u} 
\label{eq:00-78}
\end{equation}
associated with the change in mass, i.e., $\vec{F} + \vec{F}^{\rm mass}_u \to \vec{F}$.

\vskip 0.5cm
Note that the term $\dot{m}\, \vec{u}$ is a velocity-dependent force which is (inertial) frame-dependent. This is different from velocity-dependent forces like friction, where the dependence is on the velocity relative to the system. It arises directly from the considerations based on  momentum conservation theorem (see, e.g., Ref.~\cite{Sommerfeld:1952}) or Newton's second law (\ref{eq:1b}) for constant mass systems as shown in \ref{app:1}.
Expressing $\vec{u}$ as 
\begin{equation}
\vec{u} = \vec{v} + \vec{v}_{\rm rel} \ ,
\label{eq:00-79}
\end{equation}
where $\vec{v}_{\rm rel}$ denotes the relative velocity of the mass being transferred measured with respect to the system,  we obtain $\dot{m}\,\vec{u} = \dot{m}\, \vec{v}_{\rm rel} + \dot{m}\,\vec{v}$, 
showing that the (inertial) frame dependence arises from the last term, $\dot{m}\,\vec{v}$. 
Its presence is a serious issue at first sight. 
However, this term is exactly what is needed to cancel 
the same term contained in $\dot{\vec{p}}$ ($= m \dot{\vec{v}} + \dot{m} \vec{v}$) appearing in both 
Eq.~(\ref{eq:1a}) and (\ref{eq:2}); hence it is of no consequence to the resulting equation-of-motion and both Eq.~(\ref{eq:1a}) and (\ref{eq:2}) are invariant under Galilean transformation as they should be.

From the above observation, we conclude that Newton's second law in the form of Eq.~(\ref{eq:1a}) is valid for all systems in general,
provided we account for the term $\vec{F}^{\rm mass}_u$ given by Eq.~(\ref{eq:00-78}) in the external force.  Variable mass systems are \textit{not closed systems} and, as such, they are always subject to a non-vanishing external force $\vec{F}_u^{\rm mass}$.

\vskip 0.5cm
The reference to the 'spurious' (inertial) frame-dependent force, $\dot{m}\,\vec{v}$, can be eliminated from Eqs.~(\ref{eq:1a},\ref{eq:2}) by rewriting them in the form given by Eq.~(\ref{eq:1b}), with $\vec{F}$ being the total external force including 
\begin{equation}
\vec{F}^{\rm mass} \equiv  \dot{m}\, \vec{v}_{\rm rel} \ ,
\label{eq:00-80}
\end{equation}
instead of $\vec{F}^{mass}_u$ which enters in Eq.~(\ref{eq:1a}). 
Thus, Newton's second law in the form of Eq.~(\ref{eq:1b}) is valid also for a variable mass system, provided we include the force $\vec{F}^{\rm mass}$ given by Eq.~(\ref{eq:00-80}) into the external force, i.e., $\vec{F} + \vec{F}^{\rm mass} \to \vec{F}$.
In other words, variable mass systems can be treated in the same way as constant mass systems by considering an external \textit{dissipative} force ($\vec{F}^{mass}$),  
analogous to the usual frictional force, $\vec{F}\,^{\rm friction} = -k\,\vec{v}_{\rm rel}$.

\vskip 0.5cm
The preceding considerations reveal that Newton's second law can be written in different forms as given by Eqs.~(\ref{eq:1a},\ref{eq:1b},\ref{eq:2}) provided one takes into account the external forces appropriately:
\numparts
\begin{eqnarray}
{\rm Eq.}~(\ref{eq:1a}) : \ \  \vec{F} & = \dot{\vec{p}} \ , &\qquad (\vec{F}\ {\rm includes}\ \vec{F}^{\rm mass}_u) \ , 
\label{eq:EOMa} \\
{\rm Eq.}~(\ref{eq:1b}) : \ \ \vec{F} & = m\dot{\vec{v}} \ , &\qquad (\vec{F}\ {\rm includes}\ \vec{F}^{\rm mass}) \ , 
\label{eq:EOMb}\\
{\rm Eq.}~(\ref{eq:2}) : \ \ \vec{F} & = \dot{\vec{p}} - \dot{m}\, \vec{u} \ , &\qquad (\vec{F}\ {\rm excludes}\ \vec{F}^{\rm mass}_u) \ .
\label{eq:EOMc} 
\end{eqnarray}
\label{eq:EOM}
\endnumparts
In other words, as far as for practical application purposes, they are all equivalent forms and are valid for both constant and variable mass systems.

\vskip 0.5cm
We now turn our attention to a more conceptual issue involving Eqs.~(\ref{eq:EOMa},\ref{eq:EOMb},\ref{eq:EOMc}). 
When we study a dynamical system, we consider the problem in a way whereby the system under study does not accelerate when the net force acting on it vanishes.  Newton's second law  given by Eqs.~(\ref{eq:EOMa},\ref{eq:EOMb}), is then formulated to describe the dynamics of such a system with $\vec{F}$ appearing in those equations being the net force acting on the 
system. This scheme leads to the definition of the \textit{system under consideration} and the \textit{total external force} by the following two conditions:
\begin{itemize}

 \item[i)]
 The total external force is the net force acting on the system under consideration.
 
\item[ii)]
 A system under consideration is the one which does not accelerate when the total external force acting on it vanishes.

 \end{itemize}
The first condition is the usual definition of the external force. 
The second condition specifies what exactly constitutes a system under consideration. 
We note that condition (ii) is a redundancy for the case of constant mass systems, which is the reason why we don't usually see this condition stated explicitly. 

Much of the confusion and misinterpretation related to variable mass systems arises due to the lack of an explicit statement about what constitutes the \textit{system under consideration} in particular. We see that in Eq.~(\ref{eq:EOMc}), $\vec{F}$ cannot be the total external force according to (i) and (ii), for the system may accelerate even in its absence. But, the form given by Eq.~(\ref{eq:EOMc}) is obtained by calculating the variation in momentum using momentum conservation theorem and, consequently, $\vec{F}$ appearing in that equation should be understood as  the total external force apart from that already taken into account explicitly when calculating the variation in momentum.
On the other hand,  $\vec{F}$ in Newton's second law in the forms given by Eqs.~(\ref{eq:EOMa},\ref{eq:EOMb}) is meant to be the total external force. Then, in Eq.~(\ref{eq:EOMa}), $\vec{F}$ can never vanish for variable mass systems  (as has been pointed out already); otherwise the system accelerates in its absence. In Eq.~(\ref{eq:EOMb}), the system will not accelerate when $\vec{F}=0$ even for variable mass systems. Thus, in the form given by Eq.~(\ref{eq:EOMb}), the
total external force is free from any constraints. Furthermore, and perhaps more importantly, in contrast to Eq.~(\ref{eq:EOMa}), the alternative form given by Eq.~(\ref{eq:EOMb}) is free from the spurious frame-dependent force $\dot{m}\, \vec{v}$.  For these reasons, we may choose to adopt this latter form as the fundamental law, alternatively to Eq.~(\ref{eq:EOMa}) (or to Eq.~(\ref{eq:EOMc}) for that matter).

\vskip 0.2cm
At this point a remark is in order. It is common to find the statement in textbooks that Newton's second law (\ref{eq:1a}) is also a statement of momentum conservation of a closed system. However, as we have seen above, Eq.~(\ref{eq:1a}) (or Eq.~(\ref{eq:EOMa})) does not allow a mass varying system to be a closed system. 
In the context of that equation, to obtain momentum conservation ($\dot{\vec{p}} = 0$) for variable mass systems, one should first cancel the Galilean-invariance-violating term $\dot{m}\, \vec{v}$ on the right-hand side of that equation by the same term present in the external force $\vec{F}$, and then, set $\dot{m}\, \vec{v}_{rel} = - \dot{m}\, \vec{v}$. 
This is most clearly seen from Eq.~(\ref{eq:EOMb}) which leads to momentum conservation only when $\vec{F} = \vec{F}^{\rm mass} = \dot{m}\, \vec{v}_{rel} = - \dot{m}\, \vec{v}$, i.e., when there is an external force acting on the system.

\vskip 0.5cm
For completeness, in \ref{app:2}, we treat two special cases we encounter in problems involving variable mass systems.

\section{Lagrangian formalism for variable mass systems}

Variable mass systems can also be treated most straightforwardly in the Lagrangian formalism if one realizes that the force associated with varying mass is $\vec{F}^{\rm mass}$ given by Eq.~(\ref{eq:00-80}).  It is true that since $\vec{F}^{\rm mass}$ is (relative) velocity-dependent as the frictional force is, it cannot be represented by a potential and, thereby, it cannot be included into a Lagrangian. However, it can be included in the Lagrangian formalism as a non-conservative force. The resulting Euler-Lagrange equation is, then, given by
\begin{equation}
\frac{d}{dt}\left(\frac{\partial L}{\partial \dot{q}} \right) - \frac{\partial L}{\partial q} = {\cal F}\,^{\rm mass} \ ,
\label{eq:ELeq}
\end{equation}
where $L= m \vec{v}\,^2 / 2 - U(q) $ is the Lagrangian exactly as for the constant mass system, i.e., with the mass entering in it (both in the kinetic and potential energies) treated as constant.  The generalized force ${\cal F}\,^{\rm mass}$ associated with $\vec{F}\,^{\rm mass}$ is given by
\begin{equation}
{\cal F}^{\rm mass} = \vec{F}\,^{\rm mass} \cdot \left(\frac{\partial \vec{r}}{\partial q} \right) =
\dot{m}\, \vec{v}_{\rm rel} \cdot \left(\frac{\partial \vec{r}}{\partial q} \right) \ .
\label{eq:GenForce}
\end{equation}
The above result is in accordance with Newton's second law in the form given by Eq.~(\ref{eq:EOMb}).

\vskip 0.5cm
 Of course, if we wish, we could let the mass entering in the Lagrangian in Eq.~(\ref{eq:ELeq}) to be varying in time too, i.e., $m=m(t)$ everywhere. This is in accordance with Newton's second law in its form given by Eq.~(\ref{eq:EOMa}). 
In this case, the Euler-Lagrange equation becomes
\begin{equation}
\frac{d}{dt}\left(\frac{\partial L}{\partial \dot{q}} \right) - \frac{\partial L}{\partial q} = {\cal F}_u^{\rm mass} \ ,
\label{eq:ELeq1}
\end{equation}
where the generalized force ${\cal F}_u^{\rm mass}$ is now associated with $\vec{F}_u^{\rm mass}$ given by Eq.~(\ref{eq:00-78}), i.e.,
\begin{equation}
{\cal F}_u^{\rm mass} = \vec{F}_u^{\rm mass} \cdot \left(\frac{\partial \vec{r}}{\partial q} \right) =
\dot{m}\, \vec{u} \cdot \left(\frac{\partial \vec{r}}{\partial q} \right) \ , \qquad (\vec{u} = \vec{v} + \vec{v}_{rel}) \ .
\label{eq:GenForce1}
\end{equation}
The contribution to the generalized force due to $\vec{v}$ will be cancelled exactly by the contribution arising from the time dependence of the mass in the Lagrangian in the left-hand side of Eq.~(\ref{eq:ELeq1}).
The latter form (Eqs.~(\ref{eq:ELeq1},\ref{eq:GenForce1})) is precisely the way how the Lagrangian approach for variable mass systems has been presented in Ref.~\cite{Plastino:1992}, for example.

\section{Conclusion}

In conclusion, for practical applications, the equation for describing the variable mass systems can be written in equivalent forms as given by Eqs.~(\ref{eq:EOMa},\ref{eq:EOMb},\ref{eq:EOMc}). In particular, contrary to many claims in the literature, Newton's second law in its form given by Eq.~(\ref{eq:EOMa}) is applicable to both the constant and variable mass systems in general. In addition, we have also shown that Newton's second law in the form given by Eq.~(\ref{eq:EOMb}) -- commonly regarded as applicable only to constant mass systems  -- is, in fact, valid for both the constant and variable mass systems.  Moreover, from a conceptual point of view, Eq.~(\ref{eq:EOMb}) may be considered as the fundamental law, alternatively to Eq.~(\ref{eq:EOMa}), in accordance with the definitions of the system under consideration and the (total) external force given by items (i) and (ii). 

Treating variable mass systems on same footing with constant mass systems, greatly facilitates solving problems involving such systems in practice, since all what is required is 
to consider, in Eq.~(\ref{eq:EOMb}), the dissipative external force $\vec{F}\,^{mass}$ given by Eq.~(\ref{eq:00-80}). 
From a pedagogical point of view, this offers a possibility to introduce variable mass systems even to first year physics students.

Finally, we mention that, according to Truesdel \cite{Truesdel:1960}, Euler has presented the form given by Eq.~(\ref{eq:1b}) for the fist time ``as general, explicit equation for mechanical problems of all kinds"  in a paper published in 1752, ``\textit{Discovery of a new principle of Mechanics.}" In this regard, according to Ref.~\cite{Arons:1964}, the form (\ref{eq:1b}) is not to be found in Newton's ``\textit{Principia}".  Perhaps, if Euler meant to include the variable mass systems in his ``mechanical systems of all kinds", we have just rediscovered his work of over two and a half century ago. 

\vskip 0.5cm
\textbf{Acknowledgement}
\vskip 0.1cm

The author thanks Jo\~ao Plascak and Shahab Hessabi for a careful reading of the manuscript and valuable suggestions.

\appendix

\section{ }\label{app:1}

\begin{figure}[h] 
\hspace{2cm}
\includegraphics[height=0.5\textwidth,clip=1]{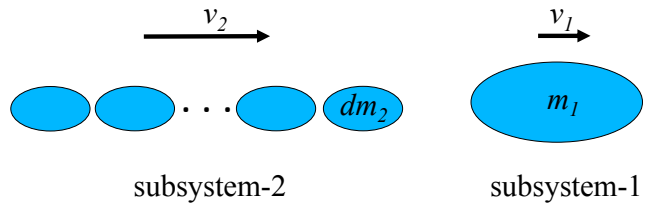} 
\vskip -4.75cm 
\caption{
A system consisting of subsystems 1 and 2. The former has mass $m_1$ and velocity $\vec{v}_1$. The latter consists of many tiny masses $d m_2$, all of which with velocity $\vec{v}_2$. The total mass of subsystem-2 is $m_2$ and its velocity $\vec{v}_2$.   
As the motion progresses, the two subsystems coalesce and subsystem-2 transfers a part of its mass, $d m_2$, to subsystem-1. 
The total system has a constant mass $M=m_1 + m_2$.}
\label{fig:fig1}
\end{figure}

Here, we re-derive Eq.~(\ref{eq:2}) -- usually obtained \cite{Sommerfeld:1952} from momentum conservation theorem -- from  Newton's second law in the form (\ref{eq:1b}) applied to a constant mass system. This  allows to better address the issue of the external force. To this end, we follow closely Ref.~\cite{deSousa:2004}.  
\footnote{The derivation can be done more sophisticatedly in terms of momentum flux by considering a velocity field associated with a continuous changing-mass distribution \cite{Thorpe:1962,Tiersten:1969} in order to also include the cases of fluid dynamics. However, for the present purpose, it suffices to treat in a simplified way. One arrives at the same conclusion either way.}
 Consider a system consisting of two subsystems. We identify one of them (subsystem-1) as body-1 and the other (subsystem-2) as body-2. The two subsystems undergo incremental mass changes, increasing or decreasing. The mass of body-1 (body-2) and its velocity -- measured with respect to an inertial reference frame -- at an arbitrary instant $t$ are $m_1$ ($m_2$) and $\vec{v}_1$ ($\vec{v}_2$), respectively. 
To be concrete, we assume that body-2 consists of many tiny masses $d m_2$, all of which with velocity $\vec{v}_2$ (see Fig.~\ref{fig:fig1}).  
%
%
As the motion progresses, the two subsystems coalesce ($|\vec{v}_2| > |\vec{v}_1|$) and that body-2 of mass $m_2$ looses a part of its mass which
immediately becomes a part of body-1. 
We should keep in mind that $m_1$ and $m_2$ are
time-dependent, but the whole system has constant total mass of $M=m_1 + m_2$, so mass is being
transferred at the rate $\dot{m}_1 = - \dot{m}_2 = d m_2 /d t$ from body-2 to body-1.

The force acting on the total system (body-1 + body-2), with constant mass $M=m_1+m_2$, follows from Eq.~(\ref{eq:1b}):
\begin{eqnarray}
\vec{F} & = M \dot{\vec{V}}_{cm} =  \dot{\vec{P}} = \dot{\vec{p}}_1 + \dot{\vec{p}}_2 \nonumber \\
& =  m_1 \dot{\vec{v}}_1  +  \dot{m}_1 \vec{v}_1 + m_2 \dot{\vec{v}}_2  - \dot{m}_1 \vec{v}_2 \ ,
\label{eq:00-74}
\end{eqnarray}
where $\dot{m}_2=-\dot{m}_1$ has been used. $\vec{V}_{cm} \equiv (m_1 \vec{v}_1 + m_2 \vec{v}_2) / M$ and $\vec{P} = \vec{p}_1 + \vec{p}_2$ stand for the center-of-mass  velocity and total momentum, respectively, of the total system.

The above force, acting on the total system, may be decomposed into the forces acting on body-1 ($\vec{F}_1$) and body-2 ($\vec{F}_2$), separately, i.e., $\vec{F} = \vec{F}_1 + \vec{F}_2$, with
\begin{eqnarray}
\vec{F}_1 & = m_1 \dot{\vec{v}}_1  + \dot{m}_1 \left(\vec{v}_1 - \vec{v}_2 \right) \ , \nonumber \\
\vec{F}_2 & = m_2 \dot{\vec{v}}_2  \ .
\label{eq:00-75}
\end{eqnarray}
We note that the above identification of the forces should be the correct one, which is in accordance with the limiting case of body-2 loosing its mass to body-1 completely. In this case, one should have only body-1 remaining in the system with mass $m_1 + m_2$. In particular, $\vec{F}_2 = 0$ when $m_2=0$. From a more physical point of view, $\dot{m}_1\,\vec{v}_2 = - \dot{m}_2\,\vec{v}_2$ is the rate at which the momentum is being transferred from body-2 to body-1 due to the transfer of the amount of mass $d m_2$ from body-2 to body-1. It is this transferred momentum that is responsible, e.g., for the recoil effect when a gun is fired. 
Of course, in this case the gun looses a part of its mass (the bullet is expelled) instead of gaining it, i.e. $\dot{m}_1 < 0$.
The recoil force is an external force. 
Equation (\ref{eq:00-75}) can be rewritten as
\begin{eqnarray}
\vec{F}_1 & = \frac{d}{dt} \left(m_1 \vec{v}_1 \right) -  \dot{m}_1 \vec{v}_2 \ , \nonumber \\
\vec{F}_2 & = \frac{d}{dt} \left(m_2 \vec{v}_2 \right) -  \dot{m}_2 \vec{v}_2   \ ,
\label{eq:00-76}
\end{eqnarray}
which reveal that the force acting upon each subsystem, where the mass is varying in time, takes the generic form as given by Eq.~(\ref{eq:2}), i.e., we reproduce the result obtained from momentum conservation theorem.
\footnote{ 
Note that in the example discussed here, the velocity of the mass being removed from body-2 is the same to that of body-2 itself ($\vec{u} = \vec{v}_2$). Of course, in general, these velocities may differ from each other as is the case for body-1 in the example considered.} 

\section{ }\label{app:2}

There are two special cases in Eq.~(\ref{eq:EOMb}) that deserve close attention:
\begin{itemize}

\item[a)]
 Mass is transferred at a relative velocity $\vec{v}_{\rm rel} = 0$ (or, equivalently, $\vec{u}=\vec{v}$ in Eqs.~(\ref{eq:EOMa},\ref{eq:EOMc})).  In this case, the external force $\vec{F}^{\rm mass}$ -- which is a part of the total external force $\vec{F}$ in Eq.~(\ref{eq:EOMb}) -- vanishes, and the equation-of-motion becomes exactly the same to the case
 of a system with constant mass, even though the mass is changing. 
 This is the situation we encounter, e.g., when a star looses a part of its mass in the form of radiation/heat with an isotropic distribution with respect to the star or in the problem of a stretched rope sliding over the edge of a table.
 
 \item[b)]
 Mass is transferred at $\vec{v}_{\rm rel}= - \vec{v}$ (or $\vec{u}=0$ in Eqs.~(\ref{eq:EOMa},\ref{eq:EOMc})),  in which case, Eq.~(\ref{eq:EOMb}) becomes $\vec{F} = m\,\dot{\vec{v}} + \dot{m}\,\vec{v} = \dot{\vec{p}}$. 
Here, at first sight, we might think that this equation violates Galilean invariance because of the presence of the term $\dot{m}\, \vec{v}$. However, this term is due to the particular value taken by $\vec{v}_{\rm rel}$, thus, it does not cause any problem. 
 This corresponds to a situation, e.g., of accretion of water drop falling through a stationary cloud or a conveyor belt on the ground transporting objects which are dropped on the belt vertically from above. 
  
\end{itemize}

\vskip 0.5cm
\textbf{References}
\vskip 0.1cm


\end{document}